\documentstyle[12pt]{article}
\begin{document}
\baselineskip=6mm
\def\Dslash{\,{\raise.15ex\hbox{/}\mkern-12mu D}}

\begin{titlepage}
\begin{flushright}
\mbox{} \\ \mbox{} \\
hep-th/9701170 \\DTP/97/10 \\
\end{flushright}
\begin{centering}
\vspace{.2in}
{\Large {\bf One-Instanton Tests of the 
Exact Results in}}
\medskip

{\Large {\bf $N=2$ Supersymmetric QCD}}\\
\vspace{.5in}
{\bf Matthew J. Slater} \\
\vspace{.1in}
Physics Department, Centre for Particle Theory  \\
University of Durham, Durham DH1 3LE, UK \\
E-mail: m.j.slater@durham.ac.uk\\
\vspace{.5in}
{\bf Abstract}\\
\end{centering}
We use the microscopic instanton calculus to determine the
one-instanton contribution to the quantum modulus
$u_3=\langle\mbox{Tr}(\phi^{3})\rangle$ in $N=2$ $SU(N_c)$
supersymmetric QCD with $N_f<2N_c$ fundamental flavors. 
This is compared with the corresponding
prediction of the hyperelliptic curves which are expected to
give exact solutions in this theory.
The results agree up to certain regular terms which appear
when $N_f\geq 2N_{c}-3$.
The curve prediction for these
terms depends upon the curve parameterization which
is generically ambiguous when $N_f\geq N_c$. In $SU(3)$ theory
our instanton calculation of the regular terms is found to
disagree with the predictions of all of the suggested curves.
For this theory we employ our results as input to improve the curve
parameterization for $N_f=3,4,5$.
\end{titlepage}

In their seminal work \cite{SW}, Seiberg and Witten
applied ideas of duality to $N=2$ supersymmetric QCD (SQCD)
with gauge group $SU(2)$ and $N_f=0,1,2,3,4$ flavors of matter
hypermultiplets, and were
able to predict exact results, valid at both strong and weak
coupling. This was achieved by identifying the quantum moduli
spaces of these models with the moduli spaces of certain
families of elliptic curves. Exact solutions for the
holomorphic prepotential describing the low energy
dynamics were then obtained via periods of a
meromorphic one-form on the curves.

This analysis has subsequently been extended to $SU(N_c)$
models with $N_c>2$ and \mbox{$N_f\leq2N_c$} fundamental
flavors
[2--5]. In the general case, the quantum moduli
space is described by a family of genus $N_c-1$ hyperelliptic
curves. These are parameterized by the gauge invariant quantum
moduli $u_n=\langle\mbox{Tr}(\phi^n)\rangle$
$(n=2,3,\ldots,N_c)$, where $\phi$ is the adjoint Higgs. The
proposed curves predict exact solutions for these objects as
well as for the holomorphic prepotential.\footnote{By virtue
of the Matone relation \cite{MAT+} (see
\cite{FT+} for instanton based derivations) between the
prepotential and $u_2$, it actually suffices to consider only
the solutions for the quantum moduli as independent
predictions of the curves when $N_f<2N_c$.}

The exact solutions can be explicitly expanded in the
semiclassical regime \cite{DKP} to give the expected one-loop
perturbative contribution plus predictions for $k$-instanton
corrections. These take the form of rational functions of the
vacuum expectation values (VEV's). Non-trivial tests of the
curves can be performed by applying the microscopic
instanton calculus to directly evaluate these non-perturbative
contributions.

In $SU(2)$, instanton calculations have been carried out
at the one-instanton \cite{FP,DKM6} and two-instanton
[11--15] levels. The results completely
agree with the curves for $N_f=0,1,2$ fundamental flavors.
However discrepancies have emerged for $N_f=3$ \cite{AHSW}
and $N_f=4$ \cite{DKM4} fundamental flavors at the
two-instanton level (as well as at the one-instanton level for
one adjoint flavor in \cite{DKM6}).
For general $N_c$, only
the singular
part of the one-instanton contribution to $u_2$ has been
calculated, in \cite{IS1,IS2}. When $N_f<2N_c-2$ there are no
additional regular terms and the result is in full agreement
with the curves. It was further claimed in \cite{IS2} that the
results of an evaluation of the regular terms that appear
when $N_f\geq4$ in the $SU(3)$ model are in conflict with the
predictions of the proposed curves.

In \cite{DKM5,HS,DKM6} it was shown that the $SU(2)$
discrepancies can be resolved in a way consistent with the
analysis of Seiberg
and Witten, by reinterpreting the parameters of the original
curves. In fact it is a generic feature of curve construction
that the curve parameterization may not be uniquely fixed when
$N_f\geq N_c$. In [2--5] possible curve
parameterizations are suggested on the basis of various
assumed criteria. The results of \cite{IS2} imply that none of
the $SU(3)$ parameterizations are correct and therefore question
the validity of these criteria.

In this letter, we continue the program of comparing the
curve predictions with the results of first-principles
instanton calculations, by evaluating the one-instanton
contribution to
the quantum modulus $u_3$ in $N=2$ SQCD with $N_c>2$ colors and
$N_f < 2N_c$ fundamental flavors. We determine the
most singular part of the answer, which for $N_f<2N_c-3$ is
the complete answer and agrees exactly with the prediction
which we extract from the curves. Our analysis also gives the
coefficients of the regular terms which arise in the 
$SU(3)$ theory when $N_f\geq 3$, and we find
disagreement with the numbers obtained from the curves, for
all of the suggested curve parameterizations. We then employ
the set of microscopic instanton
calculations in the $SU(3)$ theory to improve the
parameterizations of the $N_f=3,4,5$ curves. For $N_f=3,4$
flavors the curves should be completely fixed, so that no
discrepancies can appear at higher order instanton levels.
\medskip

The field content of $SU(N_c)$ $N=2$ SQCD is as follows. There
is an $N=2$ hypermultiplet comprised of two $N=1$ superfields,
$\Phi$ and $W_{\alpha}$, which transforms under the adjoint
representation of the gauge group. Components of the chiral
superfield $\Phi$ are a complex
scalar Higgs $\phi$ and its fermionic superpartner, the
Higgsino $\psi$. The vector superfield $W_{\alpha}$ contains
the gauge boson $A_{\mu}$ and its superpartner, the gaugino
$\lambda$. The additional $N_{f}$ matter hypermultiplets
consist of chiral superfields $Q_{f}$ and $\tilde{Q}_{f}$
($f=1,2,\ldots,N_f$), which transform under the fundamental
and its conjugate representation respectively.
The associated component fields are the squarks $q_{f}$ and
quarks $\chi_{f}$ along with their conjugate representation
counterparts, $\tilde{q}_{f}$ and $\tilde{\chi}_{f}$.

In this paper we adopt the leading-order short-distance
constrained instanton \cite{AFFLECK} approach to semiclassical
analysis as reviewed in Sections 3 and 4 of \cite{DKM1}.
The constrained Euclideanized Euler-Lagrange equations in the
short-distance region $|x| < 1/M$, where $M$ represents a
typical W-boson mass, can be solved perturbatively in the
coupling constant $g$. Only the leading-order
terms contribute to the holomorphic prepotential and quantum
moduli $u_n$. The defining equations of the instanton
configuration are consequently [11--15]
\begin{eqnarray}
\label{EoM4A}
&&\hspace{5.7cm}F_{\mu\nu}=\tilde{F}_{\mu\nu},\\
\label{EoM4ferms}
&&\hspace{2.95cm}\bar{\Dslash}\lambda=0,\quad
\bar{\Dslash}\psi=0,\quad
\bar{\Dslash}\chi=0,\quad
\bar{\Dslash}\tilde{\chi}=0,\\
\label{EoM4phi}
&&\hspace{3.1cm}D^2\phi=\sqrt{2}ig[\lambda,\psi],\quad
D^2\phi^{\dagger a}=\sqrt{2}ig\tilde{\chi}T^a \chi,\\
\label{EoM4q}
&&D^2q=\sqrt{2}ig\lambda \chi, \quad
D^2\tilde{q}=-\sqrt{2}ig\tilde{\chi}\lambda, \quad
D^2q^{\dagger}=\sqrt{2}ig\tilde{\chi}\psi, \quad
D^2\tilde{q}^{\dagger}=\sqrt{2}ig\psi \chi.
\end{eqnarray}
We use the convention $\bar{\Dslash}^{\dot{\alpha}\alpha}=
\bar{e}_{\mu}^{\dot{\alpha}\alpha}D_{\mu}$ where
$\bar{e}_{\mu}$ is the
Hermitian conjugate of $e_{\mu}=(i\vec{{\bf \sigma}},\mbox{{\bf 1}})$.
For notational
clarity we have dropped the flavor indices on the quark and
squark fields.
In the Coulomb branch of the theory, the moduli space of vacua
results from a potential term
$V(\phi)\sim \mbox{Tr}([\phi,\phi^{\dagger}]^2)$ in the
Lagrangian. Up to gauge transformations, the Higgs field
acquires the matrix of vacuum expectation values
\begin{equation}
\label{vev}
\langle\phi\rangle= \mbox{diag}(a_1,a_2,\ldots,a_{N_c}).
\end{equation}
The $a_i$ are complex parameters satisfying the constraint
$\sum_{i=1}^{N_c}a_i=0$ which ensures that
$\langle\phi\rangle$ lives in the
Lie algebra of the group $SU(N_c)$. This imposes a boundary
condition on the instanton solution for the Higgs field, since
it must approach its matrix of VEV's at large distances.

The required self-dual solution to Eq. (\ref{EoM4A}) of unit
topological charge is given by the standard $SU(2)$ pure gauge
field (BPST) instanton \cite{BPST} `minimally embedded'
in the $SU(N_{c})$ Lie algebra \cite{BCGW}. In singular gauge
this is
\begin{equation}
\label{BPSTinst}
A_{\mu} = \frac{2\rho^{2}}{g}
\frac{y_{\nu}\bar{\eta}^{a}_{\mu\nu}}{y^2(y^2+\rho^2)}
T^{a},
\end{equation}
where $y_{\mu}=(x-x_0)_{\mu}$, and $x_0$ and $\rho$ give the
location and size of the instanton respectively. We make use
of the usual 't Hooft $\eta$-symbol \cite{THOOFT} and choose a
basis of generators such that
$T^{1,2,3}_{ij}=\frac{1}{2}\sigma^{1,2,3}_{ij}$ (ie. these are
normalized Pauli matrices in the `upper left corner').

The above configuration is subject to global gauge
transformations which rotate it into the space of the
$SU(N_c)$ Lie algebra. However for the purposes of the
instanton calculation we can choose to preserve the upper
left embedding of the
BPST instanton, and perform global gauge transformations of
the matrix of VEV's (\ref{vev}) instead \cite{CORDES}.
In this case the boundary condition on the Higgs becomes
\begin{equation}
\label{bc}
\lim_{|y|\rightarrow\infty}\phi=
\Omega^{\dagger}\langle\phi\rangle\Omega
=\left(\begin{array}{cc}
A_1&A_2\\
A_3&A_4
\end{array}\right),
\end{equation}
where $\Omega\in SU(N_{c})$, and the second equality indicates
a convenient partitioning of the rotated VEV matrix; $A_1$ and
$A_4$ are $2\times2$ and $(N_{c}-2)\times(N_{c}-2)$ matrix blocks
respectively.

The action corresponding
to the leading-order instanton can be
simplified by integrating by parts and using 
Eqs. (1)--(4). We are left with  
\begin{eqnarray}
\nonumber
S_0&=&\frac{8\pi^2}{g^2}
+\int d^4x\partial_{\mu}\left\{2{\rm
Tr}(\phi^{\dagger}D_{\mu}\phi)
+(D_{\mu}q)^{\dagger}q+(D_{\mu}\tilde{q})^{\dagger}\tilde{q}
\right\} + \sqrt{2}im\int d^4x \tilde{\chi}\chi\\
\label{S0}
&&+\sqrt{2}ig\int d^4x\left(\tilde{\chi}\phi
\chi+q^{\dagger}\lambda \chi
+\tilde{q}\psi \chi\right).
\end{eqnarray}
The $\sqrt{2}$ prefactor of the quark mass term allies us
with the usual curve convention.

We now present the remaining singular gauge solutions to the
defining
equations, which we shall use to evaluate the above action.
The normalized gaugino `zero-mode' solutions are listed in
\cite{CORDES,AKMRV},
\begin{eqnarray}
\label{lambdaSC}
\lambda_{SC\alpha} & = & \frac{i\rho}{\pi}
\frac{y_{\nu}\bar{\eta}^{a}_{\mu\nu}(e_{\mu}
\bar{\xi}_{SC})_{\alpha}}{(y^2+\rho^2)^{2}}T^{a},\\
\label{lambdaSS}
\lambda_{SS\alpha} & = & \frac{\sqrt{2}\rho^{2}}{\pi}
\frac{y_{\nu}y_{\mu}\bar{\eta}^{a}_{\lambda\nu}
\eta^{b}_{\lambda\mu}(\sigma^{b}\xi_{SS})_{\alpha}}{y^{2}
(y^2+\rho^2)^{2}}T^{a},\\
\label{lambdaM}
(\lambda_{M\alpha})_{ij} & = & -\frac{\rho}{\sqrt{2}\pi}
\frac{y_{\mu}(e_{\mu}\epsilon)_{\alpha i}}
{\sqrt{y^{2}}(y^2+\rho^2)^{3/2}}\xi_{Mj}\quad \quad
(j=3,4,\ldots,N_c),\\
\label{lambdaN}
(\lambda_{N\alpha})_{ij} & = & \frac{\rho}{\sqrt{2}\pi}
\frac{y_{\mu}(e_{\mu})_{\alpha j}}
{\sqrt{y^{2}}(y^2+\rho^2)^{3/2}}\xi_{Ni}\quad \quad \quad
(i=3,4,\ldots,N_c).
\end{eqnarray}
Here $\epsilon$ is the antisymmetric tensor satisfying
$\epsilon^{12}=1$. In addition to the two
`superconformal' (SC) and two `supersymmetric' (SS) modes there
are an additional $2(N_{c}-2)$ modes
which we have chosen to partition such that the `M' modes live
in the
upper right and the `N' modes live in the lower left parts
of the matrix representation of the $SU(N_{c})$ Lie algebra. 
The analogous solutions for $\psi$ are obtained by switching
the
Grassmannian collective coordinates $\xi\rightarrow\zeta$.

The normalized solution for a quark flavor is \cite{CORDES,AKMRV}
\begin{equation}
\label{chi}
\chi_{\alpha i} = -\frac{\rho}{\pi}
\frac{y_{\mu}(e_{\mu}\epsilon)_{\alpha i}}
{\sqrt{y^2}(y^2+\rho^2)^{3/2}} \eta.
\end{equation}
The conjugate quark solution satisfies
$\tilde{\chi}_{\alpha i}=\epsilon^{ij}\chi_{\alpha j}$
provided
we exchange the Grassmannian collective coordinate
$\eta\rightarrow\tilde{\eta}$.

Turning to the scalar fields, we separate the solution for
$\phi$ into a part satisfying the homogeneous equation,
$\phi_h$, and
a particular solution $\phi_p$ which arises in the presence of
the Yukawa source term.
The homogeneous solution was found in \cite{IS1} to be
\begin{equation}
\label{phih}
\phi_h = \left(\begin{array}{cc}
\frac{y^2}{y^2+\rho^2}A_{1({\rm tl})} + \frac{1}{2} \mbox{Tr}(A_{1})
I_2 & \sqrt{\frac{y^2}{y^2+\rho^2}}A_{2} \\
\sqrt{\frac{y^2}{y^2+\rho^2}}A_{3} & A_{4}
\end{array} \right),
\end{equation}
where $A_{1({\rm tl})}=A_{1}-\frac{1}{2}\mbox{Tr}(A_{1})I_2$ and
$I_2$ is the
$2\times2$ identity matrix. This solution manifestly satisfies
the boundary condition (\ref{bc}). 

Linearity enables $\phi_p$ to be decomposed further. If we
define $\phi_{A/B}$ as the particular solution with fermionic
modes $\lambda_A$ and $\psi_B$ inserted into the source term,
then
\begin{equation}
\label{phip}
\phi_{p} = \sum_{A,B=SC,SS,M,N} \phi_{A/B}.
\end{equation}
We obtain the following list of independent solutions which
enter the right hand side of this equation:
\begin{eqnarray}
\label{sc/sc}
\phi_{SC/SC}&=&\frac{ig}{4\sqrt{2}\pi^2}
\frac{y^2(\bar{\xi}_{SC}\epsilon\sigma^{a}\bar{\zeta}_{SC})}
{(y^2+\rho^2)^{2}}T^{a},\\
\label{sc/ss}
\phi_{SC/SS}&=&-\frac{g\rho}{4\pi^2}
\frac{y_{\mu}\bar{\eta}^{a}_{\nu\mu}
(\bar{\xi}_{SC}\epsilon\bar{e}_{\nu}\zeta_{SS})}
{(y^2+\rho^2)^2}T^{a},\\
\label{ss/ss}
\phi_{SS/SS}&=&-\frac{ig\rho^2}{2\sqrt{2}\pi^2}
\frac{y_{\nu}y_{\mu}\bar{\eta}^{a}_{\lambda\nu}
\eta^{b}_{\lambda\mu}
(\xi_{SS}\epsilon\sigma^b\zeta_{SS})}
{y^2(y^2+\rho^2)^2}T^{a},\\
\label{sc/m}
(\phi_{SC/M})_{ij}&=&\frac{ig}{8\pi^2}
\frac{\sqrt{y^2}}{(y^2+\rho^2)^{3/2}}\bar{\xi}_{SC}^{i}\zeta_{
Mj},\\
\label{sc/n}
(\phi_{SC/N})_{ij}&=&\frac{ig}{8\pi^2}
\frac{\sqrt{y^2}}{(y^2+\rho^2)^{3/2}}
\zeta_{Ni}(\epsilon\bar{\xi}_{SC})_{j},\\
\label{ss/m}
(\phi_{SS/M})_{ij}&=&-\frac{ig\rho}{4\sqrt{2}\pi^2}
\frac{y_{\mu}}{\sqrt{y^2}(y^2+\rho^2)^{3/2}}
(\bar{e}_{\mu}\xi_{SS})^{i}\zeta_{Mj},\\
\label{ss/n}
(\phi_{SS/N})_{ij}&=&-\frac{ig\rho}{4\sqrt{2}\pi^2}
\frac{y_{\mu}}{\sqrt{y^2}(y^2+\rho^2)^{3/2}}
\zeta_{Ni}(\epsilon\bar{e}_{\mu}\xi_{SS})^{j},\\
\label{m/n}
(\phi_{M/N})_{ij}&=&\frac{ig}{8\sqrt{2}\pi^2}
\frac{1}{(y^2+\rho^2)}\left\{
\delta_{ij}\delta_{i,j\leq2}\sum_{k=3}^{N_c}\zeta_{Nk}\xi_{Mk}
-2\zeta_{Ni}\xi_{Mj}\delta_{i,j\geq 3}
\right\},\\
\hspace{-2cm} &&\phi_{M/M}=\phi_{N/N}=0.
\end{eqnarray}
The solution for $\phi_{A/B}$ is deduced from the solution for
$\phi_{B/A}$
by changing the sign and making the exchange 
$\xi\leftrightarrow\zeta$.

The conjugate Higgs also consists of a homogeneous and a
particular solution. The homogeneous solution is simply the
Hermitian conjugate of (\ref{phih}) whilst the particular
solution is
\begin{equation}
(\phi^{\dagger}_{p})_{ij}=\frac{ig}{4\sqrt{2}\pi^2}
\frac{1}{(y^2+\rho^2)}\left\{
\left(\frac{N_c-2}{2N_c}\right)\delta_{ij}\delta_{i,j\leq2}-
\frac{1}{N_c}\delta_{ij}\delta_{i,j\geq3}\right\}
\tilde{\eta}\eta.
\end{equation}

Finally, each squark solution is a sum of particular
solutions \cite{AKMRV},
\begin{eqnarray}
\label{qsc}
q_{SCi}&=&\frac{ig}{4\sqrt{2}\pi^2}\frac{\sqrt{y^2}}{(y^2+
\rho^2)^{3/2}}
\bar{\xi}_{SC}^{i}\eta,\\
\label{qss}
q_{SSi}&=&-\frac{ig\rho}{4\pi^2}\frac{y_{\mu}}{\sqrt{y^2}(y^2+
\rho^2)^{3/2}}
(\bar{e}_{\mu}\xi_{SS})^i\eta,\\
\label{qn}
q_{Ni}&=&\frac{ig}{4\pi^2}\frac{1}{(y^2+\rho^2)}\xi_{Ni}
\eta,
\end{eqnarray}
where $q_{A}$ represents the solution with
$\lambda_A$ inserted in the source term. The solutions for
$q^{\dagger}$ and the conjugate representation squarks may be
obtained by straightforward manipulations of these
configurations.

By plugging the above solutions into Eq. (\ref{S0}) we are
immediately able to evaluate the leading-order instanton
action. Ignoring supersymmetric zero-modes which are not
lifted and give no contribution,
we find
\begin{eqnarray}
\label{SH}
&&2\int d^4x\partial_{\mu}{\rm Tr}(\phi^{\dagger}D_{\mu}\phi)=
8\pi^2\rho^2F+g(\bar{\zeta}_{SC},\zeta_M,\zeta_N)
M(\bar{\xi}_{SC},\xi_M,\xi_N)^t,\\
&&\hspace{0.5cm}\int
d^4x\partial_{\mu}((D_{\mu}q)^{\dagger}q)=\int
d^4x\partial_{\mu}((D_{\mu}\tilde{q})^{\dagger}\tilde{q})=0,\\
&&\hspace{1cm}\sqrt{2}im\int d^4x\tilde{\chi}\chi=-i\sqrt{2}m\tilde{\eta}\eta,\\
&&\hspace{0.9cm}\sqrt{2}ig\int d^4x\tilde{\chi}\phi \chi =
-\frac{ig}{\sqrt{2}}\mbox{Tr}(A_{1})\tilde{\eta}\eta
-\frac{g^2}{24\pi^2\rho^2}
\sum_{k=3}^{N_c}(\xi_{Mk}\zeta_{Nk}+\xi_{Nk}\zeta_{Mk})
\tilde{\eta}\eta,\\
&&\hspace{-0.9cm}\sqrt{2}ig\int d^4x\left(q^{\dagger}\lambda
\chi+\tilde{q}\psi
\chi\right)=
-\frac{g^2}{12\pi^2\rho^2}\sum_{k=3}^{N_c}(\xi_{Mk}\zeta_{Nk}+\xi_{Nk}
\zeta_{Mk})\tilde{\eta}\eta.
\end{eqnarray}
In Eq. (\ref{SH}) $F$ and $M$ are the same as in
\cite{IS1}, namely
\begin{equation}
\label{F}
F={\rm Tr}(A^{\dagger}_{1({\rm tl})}A_{1({\rm tl})}
+\frac{1}{2}(A_{3}A^{\dagger}_{2}+A_{2}A^{\dagger}_{3})),
\end{equation}
and
\begin{equation}
\label{M}
M=i\left(\begin{array}{ccc}
\sqrt{2}\epsilon A^{\dagger}_{1({\rm tl})} & (A^{\dagger}_{2})^t
&
\epsilon A^{\dagger}_{3}\\ A^{\dagger}_{2} &
0 & -\frac{{\rm Tr}A^{\dagger}_{1}}{\sqrt{2}} I_{N_{c}-2}
+ \sqrt{2}A^{\dagger}_{4} \\
(\epsilon A^{\dagger}_{3})^{t} &
-\frac{{\rm Tr}A^{\dagger}_{1}}{\sqrt{2}} I_{N_{c}-2}
+\sqrt{2}(A^{\dagger}_{4})^t & 0
\end{array}\right),
\end{equation}
where $I_{N_c-2}$ is the $(N_c-2)\times (N_c-2)$ identity
matrix. The leading-order instanton action obtained by
substituting Eqs. (29)--(33) into Eq. (\ref{S0})
is in full agreement with the calculation of Ito and Sasakura
\cite{IS2}.
These authors treated
the Yukawa terms in the action perturbatively, whereas we
explicitly included
these terms as sources in the defining equations
(\ref{EoM4phi}) and (\ref{EoM4q}) in the spirit of [11--14].

Our next consideration is the measure associated with
integration over the collective coordinates which appear as
free parameters in the instanton solutions.
For the 1-instanton situation the relevant Jacobian factors
are well known \cite{THOOFT,B,AKMRV} and combine to give the
measure 
\begin{equation}
\int d\omega=2^{10}\pi^{2N_c+2}\mu^{2N_c-N_f}g^{-4N_c}
\int d\Omega\int_{0}^{\infty}\rho^{4N_c-5}d\rho\int
d^{2N_c}\zeta d^{2N_c}\xi
\int d^4x_0\int d^{N_f}\eta d^{N_f}\tilde{\eta}.
\end{equation}
Since the is action invariant under the subgroup
$H=U(1)\times SU(2)\times SU(N_c-2)$,
the $d\Omega$ integration is to be taken over the group
submanifold $SU(N_c)/H$. The
Pauli-Villars regularization mass $\mu$
is eliminated by defining the RG-invariant dynamical scale
\begin{equation}
\Lambda_{PV}^{2N_c-N_f}=\mu^{2N_c-N_f}e^{-\frac{8\pi^2}{g^2}}
\end{equation}
It is convenient to switch from the Pauli-Villars scale to
the dynamical scale used in the hyperelliptic curves.
In \cite{IS2} it was shown using renormalization group matching
arguments that the two scales are related by
\begin{equation}
\Lambda^{2N_c-N_f} =
2^{2-N_c+N_f/2}i^{N_f}\Lambda_{PV}^{2N_c-N_f}.
\end{equation}

In supersymmetric theories an important simplification of the
instanton
calculus occurs in connection with the functional integration
of the quadratic
quantum fluctuations about the instanton action. Namely the
resulting
determinant factors due to fermionic and bosonic field
fluctuations exactly
cancel each other in the background gauge \cite{DD}. We are
therefore now in a position to write down the 1-instanton
contribution to $u_{n}$. After assembling the relevant factors
and
performing the integration over the quark zero-modes we have
\begin{eqnarray}
\nonumber
u_{n}^{1I}&=&\pi^{2+2N_c}\Lambda^{2N_c-N_f}
\sum_{p=0}^{N_f}2^{8+N_c-N_f+p}g^{-4N_c+N_f-p}
\int d\tilde{\omega} \int d^{2}\zeta_{SS} d^{2}\xi_{SS} \int d^4x_0
\mbox{Tr}(\phi^n)\\
&&\times t_p \left(\mbox{Tr}(A_{1})-
\frac{ig}{4\sqrt{2}\pi^2\rho^2}\sum_{k=3}^{N_c}(\xi_{Mk}\zeta_{Nk}+\xi
_{Nk}\zeta_{Mk})
\right)^{N_f-p} \exp(-S_H),
\label{un}
\end{eqnarray}
where the $x_0$ and SS mode integrations have been separated
from $d\omega$, leaving 
\begin{equation}
\int d\tilde{\omega}=\int
d\Omega\int_{0}^{\infty}\rho^{4N_c-5}d\rho\int
d^{2N_c-2}\zeta d^{2N_c-2}\xi.
\end{equation}
$S_H$ is just the contribution of the Higgs kinetic term to
the action as given by (\ref{SH}) and the $t_p$ are symmetric polynomials
in the quark masses,
\begin{equation}
t_p = \sum_{i_1<i_2<\cdots <i_p}^{N_f} m_{i_1}m_{i_2}\ldots m_{i_p}.
\end{equation}

The Higgs field insertions into the integrand are to be
evaluated using the short-distance configurations listed
above.\footnote{This is to be
contrasted with the case where a Green's function such as
$\langle\bar{\psi}(x_1)\bar{\psi}(x_2)\bar{\lambda}(x_3)
\bar{\lambda}(x_4)\rangle$ is to be evaluated. Here it is the
limit $|x_i-x_j|\rightarrow\infty$ which is important (see eg.
\cite{FP,DKM1}) and
long-distance instanton solutions must be obtained.}
These insertions saturate the integration over the collective
coordinates corresponding to the exact supersymmetric
zero-modes. It follows that only the part of
$\mbox{Tr}(\phi^n)$ which contains precisely
four SS Grassmann variables can give a non-zero
contribution.
When $n=2$ this is just $\mbox{Tr}(\phi_{SS/SS}^2)$ and
using Eq. (\ref{ss/ss}) we can perform the integration
of the field operator over $x_0$ and the SS modes,
\begin{equation}
\label{trphi2}
\int d^{2}\zeta_{SS} d^{2}\xi_{SS} \int d^4x_0
\mbox{Tr}(\phi^2)=-\frac{g^2}{2^4\pi^2}.
\end{equation}

In \cite{IS1,IS2}, the authors considered the integral
expression (\ref{un}) when $n=2$. Since the group integration
was not generally tractable they studied the particular case
of two VEV's being infinitesimally close. In this limit they
found that the group integration linearized and could be
carried out. Their answer exhibited a singularity
structure associated with the infra-red divergence caused by
the restoration of a non-Abelian subgroup when any two VEV's
coincide. Taking this to
represent the only instance where the instanton integration
diverges, and by
considerations of dimensional analysis, gauge invariance and
holomorphy, Ito
and Sasakura deduced the full result
\begin{equation}
\label{u21I}
u_2^{1I}=\frac{\Lambda^{2N_c-N_f}}{2}\sum_{p=0}^{N_f}t_p\left(
\sum_{k=1}^{N_c}\frac{a_k^{N_f-p}}{\prod_{l\neq
k}^{N_c}(a_l-a_k)^2}
+\alpha_{N_c}\delta_{N_f-p,2N_c-2}+\beta_{N_c}\delta_{N_f-p,2N
_c}
\sum_{k=1}^{N_c}a_k^2\right).
\end{equation}
The analysis fails to determine the constant coefficients of
the regular terms, $\alpha_{N_c}$ and $\beta_{N_c}$. However
in the specific case of $SU(3)$ it was claimed in \cite{IS2}
that the
integral expression for $u_2^{1I}$ may be directly computed
and gives $(\alpha_{3},\beta_{3})=(-3/8,-15/64)$. For a range
of input values for the VEV's we have numerically verified
these results.

Now we use our explicit solutions for $\phi$ to evaluate
$u_3^{1I}$ along similar lines. This will provide the only
remaining
independent test for the $SU(3)$ curves at the 1-instanton
level.
For insertion into the integrand, we require the part of
$\mbox{Tr}(\phi^3)$ which has the necessary quadrilinear
dependence on the SS Grassmannian variables.
This is
\begin{equation}
3\mbox{Tr}\left\{\phi_{SS/SS}^2\left(\phi_h+\sum_{A,B\neq
SS}\phi_{A/B}\right)\right\}
+3\mbox{Tr}\left\{\phi_{SS/SS}\left(\sum_{A\neq SS}\phi_{A/SS}
+\sum_{B\neq SS}\phi_{SS/B}\right)^2\right\}.
\end{equation}
Since $\phi_{SS/SS}^2$ is proportional to the $2\times2$
identity matrix in the upper left block of the matrix
representation,
the first term reduces to two distinct non-zero components,
\begin{eqnarray}
\label{ss^2a}
&&\hspace{0.8cm}3\mbox{Tr}(\phi_{SS/SS}^2\phi_h)=\frac{3}{2}
\mbox{Tr}(\phi_{SS/SS}^2)\mbox{Tr}(A_1),\\
\label{ss^2b}
&&\hspace{-1.5cm}3\mbox{Tr}(\phi_{SS/SS}^2(\phi_{M/N}+\phi_{N/M
}))=
-\frac{3ig}{8\sqrt{2}\pi^2}
\frac{\mbox{Tr}(\phi_{SS/SS}^2)}{y^2+\rho^2}
\sum_{k=3}^{N_c}(\xi_{Mk}\zeta_{Nk}+\xi_{Nk}\zeta_{Mk}).
\end{eqnarray}
The second term simplifies because $\phi_{SS/SS}$
is composed of Pauli matrices living in the upper left
corner of the matrix
representation. Closer inspection shows that the only
contributing component is
\begin{equation}
\label{ss^1}
3\mbox{Tr}(\phi_{SS/SS}(\phi_{M/SS}\phi_{SS/N}
+\phi_{SS/M}\phi_{N/SS}))
=-\frac{3ig}{8\sqrt{2}\pi^2}\frac{\mbox{Tr}(\phi_{SS/SS}^2)}
{y^2+\rho^2}\sum_{k=3}^{N_c}(\xi_{Mk}\zeta_{Nk}+\xi_{Nk}\zeta_{Mk}).
\end{equation}
Upon integrating over $x_0$ and the SS modes, we get
\begin{equation}
\label{trphi3}
\int d^{2}\zeta_{SS} d^{2}\xi_{SS} \int d^4x_0
\mbox{Tr}(\phi^3)=
\frac{3}{2}\left(-\frac{g^2}{2^4\pi^2}\right)
\left(\mbox{Tr}(A_1)
-\frac{ig}{4\sqrt{2}\pi^2\rho^2}\sum_{k=3}^{N_c}(\xi_{Mk}\zeta_{Nk}
+\xi_{Nk}\zeta_{Mk})\right).
\end{equation}
The first factor in brackets is just the corresponding result
(\ref{trphi2})
for the $\mbox{Tr}(\phi^2)$ insertion whilst the second factor
precisely matches the
part of the instanton action which is pulled down by the
integration over the quark Grassmannians.

This is a remarkable result since
it allows us to immediately determine $u_3^{1I}$ from knowledge
of $u_2^{1I}$. Using
Eq. (\ref{un}) and Eq. (\ref{u21I}) and after accounting for
a rescaling of the Higgs field, we find that for $N_f<2N_c$,
\begin{equation}
\label{u31I}
u_3^{1I}=\frac{3\Lambda^{2N_c-N_f}}{2}\sum_{p=0}^{N_f}t_p\left(
\sum_{k=1}^{N_c}\frac{a_k^{N_f-p+1}}{\prod_{l\neq
k}^{N_c}(a_l-a_k)^2}
+\tilde{\alpha}_{N_c}\delta_{N_f-p,2N_c-3}+\tilde{\beta}_{N_c}
\delta_{N_f-p,2N_c-1}
\sum_{k=1}^{N_c}a_k^2\right),
\end{equation} 
where $(\tilde{\alpha}_{N_c},\tilde{\beta}_{N_c})\equiv
(\alpha_{N_c},\beta_{N_c})$. This equation is the main result
of this letter, and constitutes a non-trivial independent
prediction of the microscopic instanton calculus.

We now consider the exact results predicted by the
hyperelliptic curves
which have been proposed for $N_f<2N_c$ in
[2--5]. By making use of the freedom to shift the
$x$-variable, we can write all of the suggested curves in
the following form,
\begin{equation}
\label{curve}
y^2=P(x)^2-Q(x),
\end{equation}
where
\begin{equation}
\label{P&Q}
Q(x)=\Lambda^{2N_c-N_f}\sum_{p=0}^{N_f}t_px^{N_f-p}\quad
\mbox{and} \quad
P(x)=\prod_{k=1}^{N_c}(x-e_k)+\Lambda^{2N_c-N_f}T(x).
\end{equation}
The moduli space parameters $e_k$ satisfy $\sum_{k=1}^{N_c} e_k=0$
and are related to the moduli of the physical theory
through the formula
\begin{equation}
\label{undef}
u_n=\sum_{k=1}^{N_c}e_k^n.
\end{equation}
The function $T(x)$ satisfies
\begin{equation}
\label{T}
T(x)=\sum_{p=0}^{N_f}t_pT^{(N_f-p-N_c)}(x) \delta_{N_f-p\geq N_c},
\end{equation}
where the $T^{(N_f-p-N_c)}(x)$ are
polynomials of degree $(N_f-p-N_c)$ in $x$, with possible dependence
on the dynamical scale and also on the moduli space parameters.

It is apparent that the considerations used in constructing the curves
are insufficient to uniquely determine the function $T(x)$.
\footnote{Excepting the requirement that for $N_f=2N_c-1$
theories the $x^{N_c-1}$ term in $T^{(N_c-1)}(x)$ has coefficient
$\frac{1}{4}$ (this ensures that the meromorphic
one-form has no residue at infinity when the bare masses
are zero).}
Nonetheless the curve prediction for the
holomorphic prepotential in terms of the $a_k$ is independent
of this function \cite{DKP}.
In this respect $T(x)$ represents a superfluous degree
of freedom in the curve parameterization. However the curve
predictions for the quantum moduli $u_n$ are certainly
affected by $T(x)$. For the physical correspondence to be
complete, a definite form for $T(x)$ must exist which specifies
curves whose predictions consistently agree with the results
of instanton calculus. Furthermore, the authors of \cite{HO},
\cite{APS}, \cite{MN} and \cite{KP} all use different criteria to
propose definite forms for $T(x)$. The validity of these
criteria is open to testing by instanton calculations of the
$u_n$.

Solutions for the $u_n$ are obtained from the curves through
the periods
\begin{equation}
\label{perint}
a_k=\frac{1}{2\pi
i}\oint_{A_k}\frac{x(P'-\frac{PQ'}{2Q})}{y}dx,
\end{equation}
where the $A_k$ are a set of one-cycles enclosing branch cuts
of the curves.
These integrals can be expanded in powers of
$\Lambda^{2N_c-N_f}$ in the semiclassical regime, and the
result is \cite{DKP}
\begin{equation}
a_k=e_k+\sum_{m,n\geq 0;m+n\neq 0}
\frac{(-1)^n(\Lambda^{2N_c-N_f})^{m+n}}{(m!)^2n!2^{2m}}
\frac{\partial^{2m+n-1}}{\partial e_k}(S_k(e)^m R_k(e)^n),
\end{equation}
where
\begin{equation}
\label{S&R}
S_k(e)=\frac{\sum_{p=0}^{N_f}t_pe_k^{N_f-p}}{\prod_{l\neq
k}^{N_c}(e_k-e_l)^2} \quad
\mbox{and} \quad
R_k(e)=\frac{T(e_k)}{\prod_{l\neq k}^{N_c}(e_k-e_l)}.
\end{equation}

At the 1-instanton level it is a simple matter to invert this
series and
use the defining expression (\ref{undef}) to get the curve
prediction for
$u_n^{1I}$. The answer may be written in the form
\begin{equation}
\label{un1Icurve}
u_n^{1I}=\frac{n(n-1)\Lambda^{2N_c-N_f}}{4}
\sum_{p=0}^{N_f}t_p\left( \sum_{k=1}^{N_c}
\frac{a_k^{N_f-p+n-2}}{\prod_{l\neq k}^{N_c}(a_k-a_l)^2}
+\frac{1}{n-1}r_n^{(N_f-p)}\right),
\end{equation}
where $r_n^{(N_f-p)}$ is a regular function of the VEV's given
by
\begin{eqnarray}
\nonumber
\label{reg}
r_n^{(N_f-p)}&=&\sum_{k=1}^{N_c}
\frac{a_k^{N_f-p+n-2}}{\prod_{j\neq k}^{N_c}(a_k-a_j)^2}
\left(2a_k\sum_{l\neq k}^{N_c}\frac{1}{(a_k-a_l)} - (N_f-p+n-1)\right)\\
&&+4\delta_{N_f-p\geq N_c}\sum_{k=1}^{N_c}
\frac{a_k^{n-1}T^{(N_f-p-N_c)}(a_k)|_{\Lambda=0}}
{\prod_{j\neq k}^{N_c}(a_k-a_j)}.
\end{eqnarray}
The non-singular nature of $r_n^{(N_f-p)}$ can be verified by
expanding it in powers of the separation
between two VEV's.

We observe that when $N_f-p<2N_c-n$, the
regular function $r_n^{(N_f-p)}$ vanishes and the full answer
is unambiguously given by the singular term in Eq. (\ref{un1Icurve}).
So when $n=3$ we find complete agreement with the prediction of the
instanton analysis, Eq. (\ref{u31I}), for $N_f<2N_c-3$. 
For $n=2$ we confirm the similar observation made in \cite{IS2}, ie. the
agreement of all the proposed curves with the instanton prediction
(\ref{u21I}) when $N_f<2N_c-2$.
When $N_f\geq 2N_c-3$, the regular functions $r_2^{(2N_c-2)}$, $r_3^{(2N_c-3)}$
and $r_3^{(2N_c-1)}$ simplify to give the expected regular terms of
Eqs. (\ref{u21I}) and (\ref{u31I}), but with multiplying constants which
depend on the function $T(x)$. In Table~1 we summarize the curve
predictions for the coefficients $\alpha_3$, $\tilde{\alpha}_3$ and
$\tilde{\beta}_3$ pertinent to the $SU(3)$ theory with $N_f<6$ flavors,
according to the various suggestions for $T(x)$
in [2--5]. Our results confirm the curve predictions for $\alpha_3$
extracted in \cite{IS2}, and we see that none of the proposed curves give the
numbers predicted by instanton calculus.

We can use Eq. (\ref{un1Icurve}) and the set of instanton
calculations in $SU(3)$ theory to fix the curve
parameterization at the one-instanton
level for $N_f=3,4,5$ fundamental flavors (see Table~1).
Dimensional considerations imply that
for $N_f=3,4$ the curves are completely fixed, so that no
discrepancies should occur at higher order instanton levels.
For $N_f=5$ however, there may corrections up to the
3-instanton level. It would be interesting if an a priori
criterion for curve construction could be found which predicts
the parameterization required by the instanton calculus.
\bigskip

\begin{table}[h]
\begin{tabular}{lllllc} 
Source of prediction & \quad & $T^{(0)}(x)$ & $T^{(1)}(x)$ & $T^{(2)}(x)$ &
$(\alpha_3,\tilde{\alpha}_3,\tilde{\beta}_3)$ \\
&&&&&\\
Ref. \cite{HO} && $\frac{1}{4}$ & $\frac{1}{4}x$
& $\frac{1}{4}x^2$ & (0,0,-1/4) \\ \vspace{0.1cm}
Refs. \cite{MN} and \cite{APS} && 0 & 0 &
$\frac{1}{4}x^2+\frac{\Lambda}{48} x
+\frac{\Lambda^2}{1728}-\frac{1}{24}u_2$ & (-1,-1/2,-1/3) \\ \vspace{0.1cm}
Ref. \cite{KP} && $\frac{1}{4}$ & $\frac{1}{4}x$ &
$\frac{1}{4}x^2+\frac{1}{8}u_2$ & (0,0,0) \\ \vspace{0.1cm}
Instanton calculus && $\frac{1}{16}$ & $\frac{5}{32}x$
& $\frac{1}{4}x^2+\frac{1}{128}u_2$ & (-3/8,-3/8,-15/64) \\ 
\end{tabular}
\caption{\protect{\small {\it Predictions for the coefficients
of the regular terms
appearing in the one-instanton contributions to the moduli $u_2$
and $u_3$ in $SU(3)$ SQCD, according to suggested forms for $T(x)$}
({\it defined by the polynomials $T^{(0)}(x)$, $T^{(1)}(x)$
and $T^{(2)}(x)$}).}}
\end{table}

It is clearly desirable to extend the $SU(3)$ analysis to the
two-instanton level to check that no further discrepancies
appear for $N_f=3,4$ and to see if further curve fixing at this
level is required for $N_f=5$. It would also be interesting
to investigate the model which has $N_f=6$ fundamental
flavors, particularly in the light of the recent discoveries
for the similar $N_f=4$ model in $SU(2)$ \cite{DKM5}. There is
also the wider problem of fully evaluating one-instanton
contributions to all the quantum moduli $u_n$, for general $N_c$.
We hope to address these issues in future work.
\bigskip

\centerline{\large {\bf Acknowledgements}}
\medskip

\noindent
I would like to thank V. V. Khoze for many invaluable discussions
and N. Dorey, J. P. Nunes and D. Tong for interesting conversations.

Receipt of a PPARC UK studentship is gratefully acknowledged.

\end{document}